# Teaching a machine to see: unsupervised image segmentation and categorisation using growing neural gas and hierarchical clustering

Alex Hocking[1]⋆ James E. Geach[2]† Neil Davey[1] & Yi Sun[1]
[1] *Computer Science & Informatics Research Centre, University of Hertfordshire, Hatfield, AL10 9AB, UK*
[2] *Centre for Astrophysics Research, Science & Technology Research Institute, University of Hertfordshire, Hatfield, AL10 9AB, UK*

3 July 2015

**ABSTRACT**
We present a novel unsupervised learning approach to automatically segment and label images in astronomical surveys. Automation of this procedure will be essential as next-generation surveys enter the petabyte scale: data volumes will exceed the capability of even large crowd-sourced analyses. We demonstrate how a growing neural gas (GNG) can be used to encode the feature space of imaging data. When coupled with a technique called hierarchical clustering, imaging data can be automatically segmented and labelled by organising nodes in the GNG. The key distinction of unsupervised learning is that these labels need not be known prior to training, rather they are determined by the algorithm itself. Importantly, after training a network can be be presented with images it has never 'seen' before and provide consistent categorisation of features. As a proof-of-concept we demonstrate application on data from the *Hubble Space Telescope* Frontier Fields: images of clusters of galaxies containing a mixture of galaxy types that would easily be recognised and classified by a human inspector. By training the algorithm using one field (Abell 2744) and applying the result to another (MACS 0416.1-2403), we show how the algorithm can cleanly separate image features that a human would associate with early and late type galaxies. We suggest that the algorithm has potential as a tool in the automatic analysis and data mining of next-generation imaging and spectral surveys, and could also find application beyond astronomy.

## 1 INTRODUCTION

Machine learning is a data analysis approach that will be vital for the efficient analysis of future astronomical surveys. Even current surveys are generating more data than is practical for humans to exhaustively examine, and the next generation of survey facilities will compound the issue as we usher-in the 'petabyte' regime of astronomical research, with data acquired at a rate of many terabytes per day. For experiments such as the Large Synoptic Survey Telescope (Ivezic et al. 2008, LSST), it will be important to rapidly and automatically analyse streams of imaging data to identify interesting transient phenomena and mining the imaging data for rare sources will yield new discoveries. Even the mundane task of automatically classifying objects such as stars and galaxies of different types is well-suited to machine learning (Lahav et al. 1995).

Machine learning techniques are already applied to astronomical imaging data, however these predominantly employ *supervised* learning. There has been a good deal of effort on developing neural networks and other techniques to improve the estimation of photometric redshifts (Firth et al. 2003; Collister & Lahav 2004; Bonfield et al. 2010; Cavuoti et al. 2012; Brescia et al. 2013). More recent applications include the use of multiple layers of convolutional neural networks to classify galaxies (Dieleman et al. 2015, Galaxy Zoo) and random forests to identify transient features in PanSTARRS imaging (Wright et al. 2015), the identification and classification of Galactic filamentary structures (Riccio et al. 2015) and the inference of stellar parameters (Miller et al. 2015). Supervised learning has the disadvantage that it requires preprocessed and labelled input data. Thus, supervised learning is limited in its potential for completely automated data analysis and genuine discovery in large data sets.

Unsupervised learning offers a solution by eliminating the need for human intervention (e.g. pre-labelling) in the learning process. Instead, unsupervised techniques use simple rules for the mapping of the 'feature space' of a given data set, ultimately encoding a representation of the topology of the data that identifies structure of interest. An obvious application is to astronomical imaging, where one might wish to identify and classify sources such as galaxies or stars; the potential for this has been recognised for over two decades (Klusch & Napiwotzki 1993; Nielsen & Odewahn 1994; Odewahn 1995), but arguably yet to be properly realised.

⋆ a.hocking3@herts.ac.uk
† j.geach@herts.ac.uk

While a human can perform such a task almost intuitively (but subjectively) by 'eyeballing' an image, this is only practical for small areas of the sky and for data with good signal-to-noise. By teaching a machine to 'see' an image in the equivalent manner, one could set it the task of 'eyeballing' the entire sky at a much greater level of detail than a human (or even group of humans) could, with the added benefit that the machine would be entirely consistent in its classification.

Unsupervised learning has found application in astronomy, particularly in the estimation of photometric redshifts (Geach 2012; Way & Klose 2012; Carrasco Kind & Brunner 2014) or object classification from photometry or spectroscopy (D'Abrusco et al. 2012; in der Au et al. 2012; Fustes et al. 2013). Yet unsupervised machine learning techniques have not been fully applied to astronomical image (pixel) data directly with a goal of performing automatic classification and labelling. Recent work by Schutter & Shamir (2015) presents computer vision techniques to identify galaxy types (see also Banerji et al. 2010). However this approach still requires an existing catalogue of galaxy images, where each image contains a single, centred galaxy.

Here we apply a combination of unsupervised machine learning algorithms directly to astronomical images, using the information content of the images themselves to derive appropriate classifications via a pixel segmentation and labelling process. As a demonstration of proof-of-concept we apply the algorithm to *Hubble Space Telescope (HST)* Frontier Fields (FF)[1] observations of two massive clusters of galaxies – fields that contain a mixture of different types of galaxy that offer an ideal test case. We use one FF as a training image and apply the trained algorithm to another 'unseen' FF image. In this trivial example the goal is to demonstrate that the algorithm can cleanly separate early and late type galaxies in the image, having simply given the algorithm simple instructions on how to 'see' the image. The Letter is organised as follows: in §2 we introduce and describe the unsupervised algorithms, in §3 we apply these to the FF data, presenting our results and commenting on the labelling methods used. We conclude in §4 with a comment on the limitations of our method and avenues for future development, as well as other applications.

## 2 THE ALGORITHM

A class of unsupervised machine learning algorithms called clustering algorithms can be used to discover structure in multidimensional data. These algorithms are used to identify regions of density in the parameter space describing the data, known as clusters. For simple data analysis the process may end here, but we can also consider the clusters as a model that can be used to identify similar structure in new unseen data. Generally, unsupervised machine learning algorithms require as input a feature matrix (FM), comprised of a large number of feature vectors sampled from the training set. The goal of the algorithm is to identify clusters within the FM.

Many unsupervised algorithms require a predefined number of clusters to describe the data (e.g. *k*-means), but without detailed knowledge of the data this is very difficult to estimate *a priori*, thus limiting the potential for discovery in large data sets. In this work we present a technique whereby the algorithm itself defines the number of clusters, overcoming this limitation. We use a combination of Growing Neural Gas (GNG, Fritzke et al. 1995, See Section 2.1) which maps the topology of a data set by fitting a graph to the data, and a variant of hierarchical clustering (Hastie et al. 2009, See Section 2.2) called agglomerative clustering to identify clusters in the GNG output. In the following sections we describe these two main components in more detail.

[1] https://archive.stsci.edu/prepds/frontier/

### 2.1 Growing Neural Gas

The GNG algorithm (Fritzke et al. 1995) identifies structure within data by creating a topological map of the FM. It does this by iteratively growing a graph to map the FM. The graph consists of nodes connected by lines called edges. Each node has a position which consists of an *n*-dimensional vector called the position vector. The dimensionality of the node position vector has the same dimensionality as the samples in the FM. The algorithm starts by creating a graph of two nodes. Each node is initialised using a random sample from the FM. The graph grows and shrinks as the input data is processed (i.e. more samples are introduced). During this process the positions of the nodes evolve: the node position vectors are updated to map the topology of the data and the graph splits to form disconnected sub graphs, each of which represents a cluster within the FM. The process continues until a stopping criterion has been met, such as a saturation value for the number of nodes within the graphs (we set this arbitrarily to 40,000), or the processing time (we set this to the equivalent of the number of processing steps that would sample the entire FM 100 times). In order to create a graph that accurately maps the input data it is common to process the input data multiple times. The learning steps of the algorithm are:

(i) *Initialization* Create a graph with two nodes. Initialise the position of each node with the vector of values from a random vector from the FM. Subsequently, samples are randomly drawn from the FM and the following set of rules applied:

(ii) *Identify the two nodes nearest to the sample vector* For each node in the graph, the distance $d$ between the sample vector **p** and the node's position vector **q** is calculated using the squared Euclidean distance (1). The two nodes ($n_0$, $n_1$) with the greatest similarity to the sample vector (i.e. the two smallest values of $d$) are identified.

$$d(\mathbf{p}, \mathbf{q})^2 = \sum_{i=1}^{n} (q_i - p_i)^2 \qquad (1)$$

(iii) *Increase the age of the graph edges connected to the nearest node $n_0$* The 'age' of each edge is incremented by one unit.

(iv) *Increase the 'error' of the nearest node $n_0$* The 'error' is simply the squared Euclidean distance between a sample vector and nodes in the GNG: if the error is high then the GNG has not yet properly mapped the parameter space containing the sample vector. In this step the squared Euclidean distance between the input sample vector and $n_0$ is added to the error of $n_0$.

(v) *Move the nearest node $n_0$* The $n_0$ position vector is subtracted from the input sample vector, multiplied by a weighting parameter $\epsilon_n = 0.0006$ and the result added to the $n_0$ position vector. This step moves the nearest node 'towards' the input sample. The $\epsilon_n$ parameter controls the size of the movement towards the input sample.

(vi) *Move connecting nodes' neighbours* Using the same process as in the previous step but using the $\epsilon_b = 0.2$ parameter to con-

trol the magnitude of the adjustment for nodes directly connected to $n_0$.

(vii) *Reset age of the $n_0$–$n_1$ edge* If an edge connects $n_0$ to $n_1$ then reset the age of the edge to zero.

(viii) *Add an edge* If there is no edge between $n_0$ and $n_1$ then add an edge to connect the two nodes.

(ix) *Remove old edges and nodes* Remove all edges with an age greater than the maximum age $A$ parameter. All nodes without edges are removed.

(x) *Add a new node to the GNG graph* Every $\lambda = 50$ samples add a new node at the midpoint between the node with the highest error and its connecting node. If multiple nodes are connected then the new node is positioned at the midpoint of the connecting nodes with the highest error. When a new node is added, the error of each node is reduced by $\alpha = 0.5$.

(xi) *Reduce all node error values* Reduce the error of each node in the GNG graph by a factor of $\beta = 0.995$.

Fritzke et al. (1995) describe the parameters mentioned above; our work uses the default values. Tests were run to identify any improvements using different sets of parameter values but no significant improvements were found. The majority of the compute time is in step (ii); various attempts have been made to reduce the time taken (Fiser et al. 2012; Mendes et al. 2013), but no optimal solution has yet been found. We increase performance by executing the nearest neighbour calculation in parallel. The set of clusters identified by the GNG represent an abstract mapping of the input data feature space as characterised by the FM. We now apply a series of reduction methods to hierarchically group these clusters, building a data structure that can be used to classify various 'levels of detail' in the input data.

**2.2 Hierarchical Clustering**

Hierarchical clustering (Hastie et al. 2009) involves a recursive process to form a hierarchical representation of a data set as a tree of clusters. One of the key benefits of HC is that it can produce uneven clusters, both in terms of their disparate sizes and separation in the parameter volume. Many unsupervised learning algorithms produce even cluster sizes which implies an assumption about the structure of the data; HC makes no such assumption. The process starts by merging pairs of data samples (*n*-dimensional vectors) from a data set to form parent clusters. The process continues recursively by performing the same merging process on pairs of parent clusters until sufficient iterations have occurred that a single cluster remains (the 'root') or, alternatively, a threshold similarity value is reached, signalling the end of the process. The identified clusters form a hierarchical representation of the input data.

This hierarchical representation can be thought of as a tree structure where the leaves represent the individual input vectors from the data set. The process starts by merging pairs of leaves, using a measure of dissimilarity to identify the most similar pair of leaves. The pair with the closest proximity are merged into a new cluster (twig) that is added to the tree as a new parent node to the pair. The process continues by merging pairs of nodes at each level until a single node remains at the root of the tree. The final tree representation contains multiple 'levels' of nodes, with each node in a level representing a cluster. Each level can be considered a level of detail in a clustered representation of the data. Our approach is to apply HC to the *output* of the GNG, further refining this representation of the input data into a cluster hierarchy that can be used to segment and classify image components.

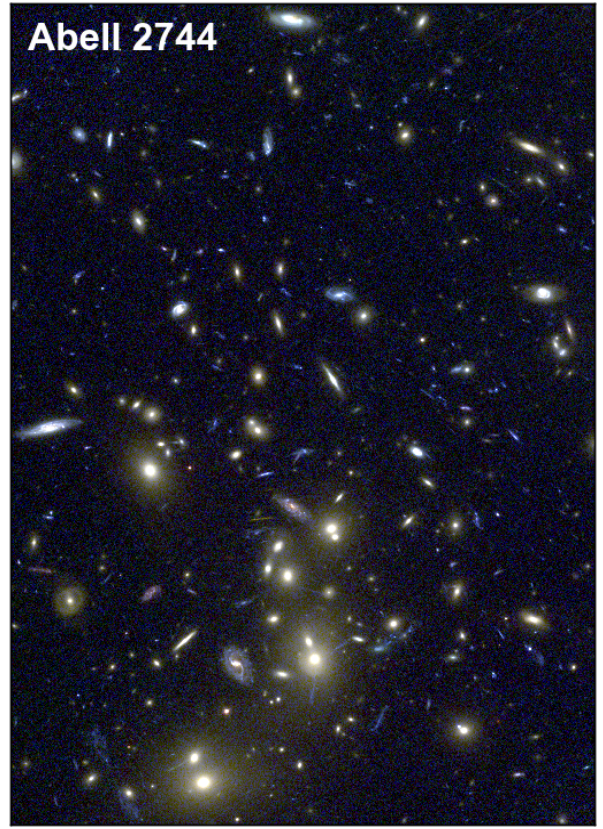

**Figure 1.** Training data for our proof-of-concept example. This is an RGB composite image of the *HST* Frontier Field Abell 2744. The red, green and blue channels correspond to the F814W, F606W and F435W filters. We chose this data set as it represents a classic example of object segregation that is well understood: the cluster dominated by red elliptical galaxies, surrounded by blue late types and gravitationally lensed features. In our proof-of-concept our goal is to demonstrate that the GNG+HC algorithm can cleanly classify these two basic classes automatically. Importantly, since the Frontier Fields target several clusters, we can test the algorithm on a different cluster, representing an image that the algorithm has not yet seen.

There are a number of approaches to measure similarity between vectors, including Euclidean distance, Pearson correlation and cosine distance. After experimenting wih these three types we found the best results were obtained using the Pearson correlation coefficient

$$r(\mathbf{p}, \mathbf{q}) = \text{cov}(\mathbf{p}, \mathbf{q})\text{var}(\mathbf{p})^{-0.5}\text{var}(\mathbf{q})^{-0.5} \qquad (2)$$

where $r$ is the Pearson correlation between $\mathbf{p}$ and $\mathbf{q}$ (the position vectors from two GNG graph nodes). Linkage concerns the process of using the similarity measure to merge clusters. We apply 'average' linkage which uses the Pearson correlation similarity measure to compare the centroids of the clusters at each level of the tree; a centroid is calculated by finding the average sample value within a cluster. After assessing the pairwise distance between all clusters in a level, clusters with the minimum linkage are merged, and the centroid of the merged cluster recalculated, ready for the next merging step as one moves up the hierarchy towards the single root.

Each node in the tree can be given a unique label and so the input data can be classified according to which node in the tree best describes it, as some desired level of detail (the trivial exam-ple is that the 'root' by definition would label all of the data). In this work we are concerned with imaging data, and the algorithm described above can be used to label individual (or groups) of pixels in an image, therefore automatically segmenting and classifying them. Consider an image containing two different types of object: provided the FM captures the difference between these objects (be it morphology, colour, intensity, etc.), then the process described above should automatically identify and label these two objects differently. In the following we demonstrate this with real astronomical data.

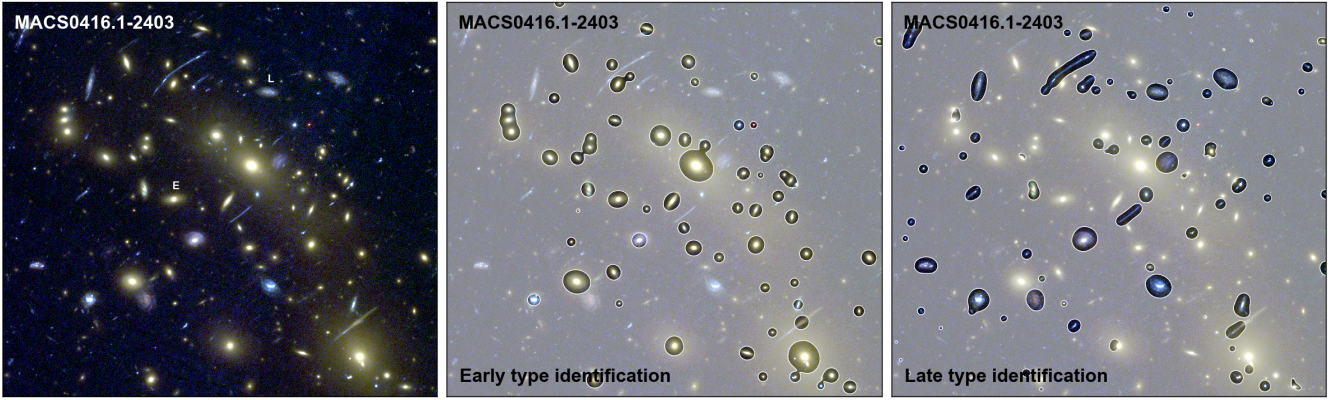

**Figure 2.** Segmented image showing classifications of features in the field of MACS 0416.1-2403, using the imaging data of Abell 2744 as a training set. The left panel shows the RGB image of the cluster comprised of the F435W, F606W and F814W bands. The feature matrix was extracted from this image in the same manner as the training set (described in §3). The central and right panels show the regions of the image classified as 'early type' and 'late type', following the label-grouping procedure described in §3. Contours show the rough boundaries of these two classifications. It is clear that the algorithm has successfully segregated objects that a human would also identify by 'eyeballing' the colour image on the left.

## 3 APPLICATION: THE HUBBLE FRONTIER FIELDS

We use deep *Hubble Space Telescope (HST)* images (F435W, F606W and F814W filters) of the strong lensing galaxy clusters Abell 2744 and MACS 0416.1-2403 to demonstrate proof-of-concept and practical application of the algorithm. Since images of clusters contain two distinct galaxy populations (namely bright early types and the numerous blue late types and background galaxies, including gravitationally lensed features), this data provides an excellent opportunity to test whether the algorithm can automatically identify and distinguish these classes of object that a human could trivially do 'by eye'.

The FM comprises a sequence of $8 \times 8$ pixel thumbnails sampled from each of the training images (the aligned F435W, F606W and F814W images of Abell 2744, Figure 1). For each thumbnail we evaluate the radially averaged power spectrum in five bins, allowing us to encode information about the pixel intensity and distribution over nine pixel scales in a manner that is rotationally independent. The power spectrum for each filter is concatenated into a single 15-element vector, giving colour information to the FM. To improve speed, during training we only consider regions of the image with pixel values in excess of $5\sigma$, where $\sigma$ is the root mean squared value of blank sky in the image[2].

The GNG algorithm (§2.1) generates over 9,000 clusters, and these are grouped according to the hierarchical scheme described in §2.2. In this demonstration we select the 60 largest clusters in the hierarchy (representing 97% of the image area) to use in a segmentation procedure, whereby pixels in new input data are labelled (classified) using their cluster ID. Having trained the network using Abell 2744, we can use the unseen data MACS 0416.1-2403 to test the algorithm (as the latter is 'unseen'). Figure 2 shows a portion of the Frontier Field data in the vicinity of the cluster core, containing elliptical galaxies, late types and lensed features. As a simple demonstration of the efficacy of this algorithm we pick two galaxies in the image: an early type and a late type (indicated E and L on the image). For ease of illustration in this case we group the labels given by the GNG+HC for each of these galaxies and gather them into a single label each. Then, through the principle of equivalence, all other labels throughout the image are relabelled accordingly. In other words, if the set of labels describing E is $\{1, 2, 4, 7\}$, then we perform the relabelling $1 \rightarrow 1, 2 \rightarrow 1, 4 \rightarrow 1, 7 \rightarrow 1$ throughout the image. An identical step is applied to the set of labels describing galaxy L for label '2'.

It is important to stress that we are manually applying this step as demonstration of proof-of-concept, since the separation of early-type and late-type galaxies is well understood in astronomy. In general one will not have this prior knowledge, and so the labels from the hierarchy can be used directly. A connective relabelling procedure as just described can be applied if necessary to reduce the total number of different classifications if desired, but this step can be automated to maintain the unsupervised nature of our approach.

Figure 2 shows the result of the relabelling step for 'early types' and 'late types' as identified by the algorithm. We have highlighted the parts of the image that were labelled as '1' or '2' using transparency and outlined those regions with contours for better visibility. It is clear that the algorithm has successfully separated early and late types (the latter including lensed features that would be immediately recognised by eye. It is important to note that some galaxies have ambiguous labels, with dual classifications. In this particular example this is due to spatially resolved spiral galaxies also being resolved by the algorithm, with bulges identified as early types (1) and the spiral structure identified as late type (2). This highlights the importance of using multiple levels of segmentation in the image classification, and the potential power in the method for automatically recognising fine structure.

We have not fully optimized the algorithm for speed (and the specific processing speed will depend on the complexity of the fea-ture matrix), however as a guide, in the example presented here the training process on the Abell 2744 imaging took 39.5 msec per pixel and the application of the trained algorithm to the new MACS 0416.1-2403 image took 1.7 msec per pixel. The work was performed on a desktop Intel Core i7-3770T 2.50GHz with 8GB RAM. These performances can clearly be dramatically improved, especially through the use of graphical processor units and optimal threading; we are currently porting parts of the code to Compute Unified Device Architecture (CUDA) language. The classification process is fully parallelisable, and the compute time for classification scales linearly with the number of pixels for a given model, making this a highly efficient algorithm to apply to large imaging data.

---
[2] Although note that in principle this data could be used during training

## 4 SUMMARY

We present a novel, efficient unsupervised machine learning algorithm that uses a combination of growing neural gas and hierarchical clustering to automatically segment and label imaging data. As a demonstration we have applied the algorithm to data from the *HST* Frontier Fields survey, showing how the algorithm can be trained using data from one field (Abell 2744) and applied to another 'unseen' data set (MACS 0416.1-2403) to successfully identify parts of the image that would be classified as 'early' and 'late' types by a human inspector. The unsupervised nature of the algorithm makes it ideal for automated processing of imaging data that contains structures that are either completely unknown (such as a rare class of galaxy) or extremely subtle (features close to the signal-to-noise limit for example). As such, it offers the potential for discovery by mining the feature space of large imaging surveys: LSST and *Euclid* are obvious target sources that will benefit greatly from automatic classification techniques. In the case of LSST, the time domain could be introduced into the feature matrix as a route to classifying transient phenomena.

There are limitations to the method that should be noted. The most significant is the choice of the feature matrix. In this work we use feature vectors that effectively encode information about colour, intensity and intensity distribution on small scales. In principle the feature vector can be arbitrarily large, but at the cost of computation time; therefore there is a balance between performance and the sophistication of the feature matrix. It is clear that the exact choice of feature matrix will have an impact on the ability of the algorithm to successfully segment and classify input data, and the optimisation of this is far from clear. It is possible that one could devise a pre-screening algorithm that actually identifies the optimal set of features to use in the training, but that is beyond the scope of this paper.

We conclude by noting that the algorithm presented here is not limited to imaging data: spectral data could also be passed through the GNG+HC, which may be relevant to current and next generation radio surveys. Indeed, the algorithm is completely general and one can envision applications beyond astronomy, in medical, satellite or security imagery and beyond.

## ACKNOWLEDGEMENTS

The authors thank N. Hine for comments on the manuscript. J.E.G. acknowledges the support of the Royal Society through a University Research Fellowship. We gratefully acknowledge the support of NVIDIA Corporation with the donation of the Tesla K40 GPU used for this research.

This paper has been typeset from a TEX/LATEX file prepared by the author.